\documentclass{svproc}
\usepackage{graphicx}
\usepackage{listings}

\begin{document}

\title{An Efficient B-tree Implementation for Memory-Constrained Embedded Systems}

\titlerunning{B-tree Implementation for Memory-Constrained Embedded Systems}
\author{Nadir Ould-Khessal\inst{1} \and Scott Fazackerley\inst{1} \and Ramon Lawrence\inst{2}}

\authorrunning{Ould-Khessal, Fazackerley, Lawrence} 
\tocauthor{Nadir Ould-Khessal, Scott Fazackerley, Ramon Lawrence}

\index{Ould-Khessal, N.}
\index{Fazackerley, S.}
\index{Lawrence, R.}

\institute{Dept. of Electronic Engineering Technology\\
Okanagan College, Kelowna, BC, Canada\\
\email{NKhessal@okanagan.bc.ca, SFazackerley@okanagan.bc.ca}
\and
Department of Computer Science, University of British Columbia \\ Kelowna, BC, Canada, V1V 2Z3\\
\email{ramon.lawrence@ubc.ca}}

\maketitle  

\begin{abstract}
Embedded devices collect and process significant amounts of data in a variety of applications including environmental monitoring, industrial automation and control, and other Internet of Things (IoT) applications. Storing data efficiently is critically important, especially when the device must perform local processing on the data. The most widely used data structure for high performance query and insert is the B-tree. However, existing implementations consume too much memory for small embedded devices and often rely on operating system support. This work presents an extremely memory efficient implementation of B-trees for embedded devices that functions on the smallest devices and does not require an operating system. Experimental results demonstrate that the B-tree implementation can run on devices with as little as 4 KB of RAM while efficiently processing thousands of records.
\end{abstract}


\section{Introduction}

Processing data on devices where it is collected reduces network transmissions, improves response time, and minimizes energy usage. Edge data processing requires the embedded sensor device store, query, and manipulate the data locally without relying on external cloud resources. The fundamental challenge is performing the data processing given that embedded devices have limited CPU and memory resources, and data processing must share resources with the critical data collection activities.

Embedded devices that collect sensed data, such as data loggers and wireless sensors, often store the data in persistent storage. The simplest storage technique is using sequential files, but such storage is extremely inefficient for querying and data processing. B-trees are a data indexing structure that provide $O(logn)$ performance for reads and writes and are widely used in database systems. The challenge is that existing B-tree implementations assume a significant amount of available memory and often use operating system features for memory and file management. On small embedded devices, such as the Arduino \cite{arduino}, the available RAM may range between 4 KB to 32 KB and an operating system may not be available.  Devices are also subject to unexpected reset and power cycling so data must not be left in SRAM and moved to persistent storage when modified to provide fault tolerance. In this environment, the B-tree implementation must minimize memory usage and handle many low-level file and memory manipulation operations itself.

This work presents a memory-optimized B-tree implementation that has high performance for embedded devices. The B-tree requires only two page buffers plus less than 100 bytes of RAM for state and temporary variables. The implementation uses no dynamic memory (i.e. malloc()) and performs all its functions with the pre-allocated buffers and state variables. This allows the B-tree to use less than 1.5 KB of memory and require only about 10 KB of flash code space. Using various memory saving techniques, this reduces the memory usage by about three times compared to other implementations.

The contributions of this work are:

\begin{itemize}
    \item A memory-efficient B-tree implementation that requires only 2 page buffers and less than 1.5 KB of RAM.
    
    \item Optimizations to reduce the number of memory buffers used during the split operation for a record insert.
    
    \item A performance evaluation on memory-constrained embedded devices that demonstrates the efficiency and practical use of the B-tree for sensor data collection and analysis.
    
\end{itemize}

The paper begins with a background on indexing on embedded devices. A description of the B-tree implementation follows, and then an experimental section demonstrates performance benefits. The paper closes with future work and conclusions.

\section{Background}

Although there are numerous indexing approaches for flash memory and solid-state drives (SSDs) \cite{indexsurvey}, there are considerably fewer research results on indexing for embedded devices. Embedded indexing has many of the same challenges as indexing for SSDs including handling the different read and write performance and the erase-before-write constraint (i.e. no overwriting or in-place writes). In addition, embedded index algorithms must minimize the memory used, as RAM is one of the most limited resources on embedded systems. Due to these RAM limitations, it is not possible to directly use an indexing system designed for flash memory/SSDs on embedded devices. Optimizations for flash indexing include performing sequential instead of random writes, using more read operations to reduce the number of writes, and exploiting hardware specific parallelism. 

There are implementations of flash-optimized B-trees that adapt to flash characteristics using one of two general approaches. The first approach is to defer writes by buffering nodes in memory instead of writing them immediately after update. Write-optimized trees \cite{betree} buffer modifications and write them out in batches to amortize the write cost and avoid small random writes. Strategies based on buffering have limited use on embedded devices as the memory available is too low to buffer enough data for a significant benefit.

The second approach logs changes to the tree rather than updating the tree immediately. The logged changes may be recorded in another area in memory/flash \cite{asbtree,bftl} or in special areas of the flash data page \cite{lsbtree,iplbtree}. Log-structured merge trees (LSM-trees) \cite{lsmtree} consist of multiple levels of B-trees with the top-level buffered in memory. LSM-tree variants offer high write performance with a trade-off of lower read performance and a large amount of memory consumed. For certain memory types, it may be possible to perform partial page flash updates in restricted cases \cite{btreeoverwrite2}. 

Indexing on servers with SSDs has different challenges than indexing on an embedded device. A small-memory embedded device has between 4 KB and 32 KB of SRAM and a processor running between 16 and 128 MHz. Example hardware includes the ATMega2560\footnote{www.microchip.com/wwwproducts/en/ATmega2560} (16 MHz, 8 KB) that are used in Arduino devices and the PIC18F57Q43\footnote{www.microchip.com/wwwproducts/en/PIC18F57Q43} (16 MHz, 8 KB). The flash storage is either raw NOR or NAND memory chips or an SD card. 

For most flash storage, the unit of read and write is a page. A page may have between 256 and 4096 bytes. Before a page is written, it must be erased. Erasing happens in units of blocks, which are consecutive pages. Due to the erase before write constraint, writing updated data to the same page address is rarely performed due to cost. Instead, the flash translation layer (FTL) writes to a new physical location, and updates the logical to physical page mapping to make this transparent to the code using the storage. When using raw memory with no FTL, the algorithm must manage physical page allocation, garbage collection, and wear leveling itself. Some NOR flash memory allow direct byte-addressable reads and writes. In this work, the focus is page-level I/O. Another requirement is that data needs to be moved to persistent storage rapidly and not left in SRAM buffers due to the increased risk of unexpected reset or power cycling. 

Indexing approaches specifically designed for embedded devices include Antelope \cite{dbeverysensor}, MicroHash \cite{microhash}, PBFilter~\cite{pbfilter}, and SBITS \cite{sbits}. Antelope is a database system for sensor devices that includes a sequential, inline file index structure that stores sorted time series data. This is a specialized index for sorted data (e.g. an in-order time series) and not a general-purpose index. MicroHash \cite{microhash} also uses a sorted time series data file, and builds an index structure supporting queries by value on top of it. The value index consists of a directory of buckets, with each bucket spanning a range of values. Each index entry stores a page and offset of the record with that value. Optimizations tried to completely fill index pages and minimize the number of index page writes.  PBFilter \cite{pbfilter} minimizes memory usage by sequentially writing the data and index structure. The index structure uses Bloom filters to summarize page contents, and bitmap indexes for handling range queries and key duplicates. SBITS \cite{sbits} uses an efficient sequential data and index structure for time series data and was shown to use less memory than MicroHash and PBFilter while maximizing insert efficiency. Linear hashing has also been implemented for embedded devices \cite{linearhash}. These embedded index techniques often index sequential time series data rather than a general data set. Notably, to our knowledge, there is no published work on a B-tree implementation for small memory embedded devices.

In summary, the existing B-tree implementations for flash memory/SSDs achieve higher performance by exploiting I/O parallelism of SSDs and the extensive memory available on servers for buffering updates. These implementations are not executable on the embedded platforms investigated as SRAM is extremely limited. This work develops a B-tree implementation that is efficient for the smallest embedded devices.

\section{B-Tree Implementation}

The primary challenge with implementing a B-tree for embedded devices is minimizing the memory usage. The absolute minimum memory required is two buffers: one write buffer for page modifications and one read buffer for reading pages. The implementation uses the write buffer for record inserts and tree modifications. Figure \ref{insert} contains pseudocode for the insert operation, and Figure \ref{split} describes the split operation.

\begin{figure}[htbp]
\begin{lstlisting}[language=Java,basicstyle=\footnotesize\ttfamily]
for each level of the tree starting at the root:
    read page into read buffer
    store page id in active path
    search page to find next child node to read
    
read leaf node into write buffer

if space available on page:
    insert record in sorted order on page
    write page to storage
else:
    perform split
\end{lstlisting}
\caption{B-tree Insert Implementation}
\label{insert}
\end{figure}

\begin{figure}[htbp]
\begin{lstlisting}[language=Java,basicstyle=\footnotesize\ttfamily]
while not above root
    if space available on page:
        insert record in sorted order on page
        write page to storage
        return
        
    mid = count / 2
    insertLoc = insertion location in node for key
    buffer record at mid in temporary record storage
    if insertLoc < mid:
        // Configure small (left) page in the split
        shift records starting at insertLoc down 1
        write insert record to location insertLoc
        update record count
        write (left) page to storage 
        
        // Configure large (right) page in the split
        write record in temp storage to start of page
        copy records after mid to start of page
        write (right) page to storage 
    else:
        // Configure small (left) page in the split
        set count of page to mid+1
        write (left) page to storage 
        
        // Configure large (right) page in the split
        copy records after mid and before insertLoc 
              to the start of the page
        write insert record at insertLoc-mid
        copy records after insertLoc to start of page
        write (right) page to storage 
        
    current node is next node on active path

create new root
add record to root
add left and right pointers to root
\end{lstlisting}
\caption{B-tree Split Implementation}
\label{split}
\end{figure}

When performing a split, the input is the record to be inserted in the full node and a left and right child pointer if the node being split is an interior node. There are two cases. If the insertion location of the insert record into the node is less than the mid point (where the split is being performed), then it is necessary to shift records down from the insert location to make room for the record to be inserted. The record is inserted, and the left page with the smaller values in the split is written to storage. Note that the page is not cleared of the records that are no longer in this page, only the count of the number of records is updated. This old record data at the end of the page is ignored. Not overwriting these records is necessary to avoid needing a second buffer. To create the right page with the larger values, the mid point record is copied to the start of the record followed by the remaining records on the page. 

The case where the insertion location is after the midpoint is similar. Updating the left page requires only updating the count and writing out the page. The right page is produced by shifting the records to the front of the existing page and inserting the new insert record in the proper sorted order.

After one iteration is completed, there is a new insertion record containing the mid point record in the split with its associated left and right child pointers. Splitting may occur all the way to the root in which case a new root node is built with the record and left and right pointers. The path from root to leaf during the insert is stored in an array when searching for the insertion leaf so that it is easy to backup the path when handling splits starting at a leaf.

\subsection{Example and Analysis}

\begin{figure}[!htb]
	\centering
	\includegraphics[width=4.5in]{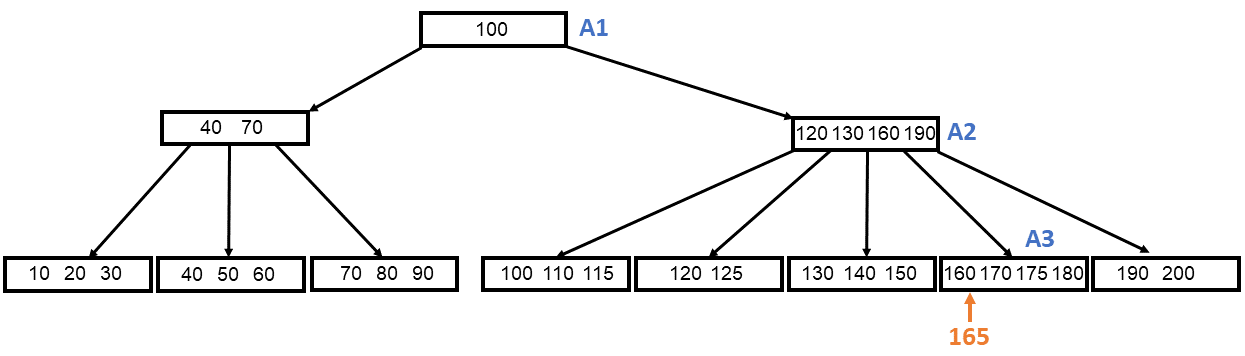}
	\caption{B-tree Before Insert of 165}
	\label{exampletree}
\end{figure}

\begin{figure}[!htb]
	\centering
	\includegraphics[width=4.5in]{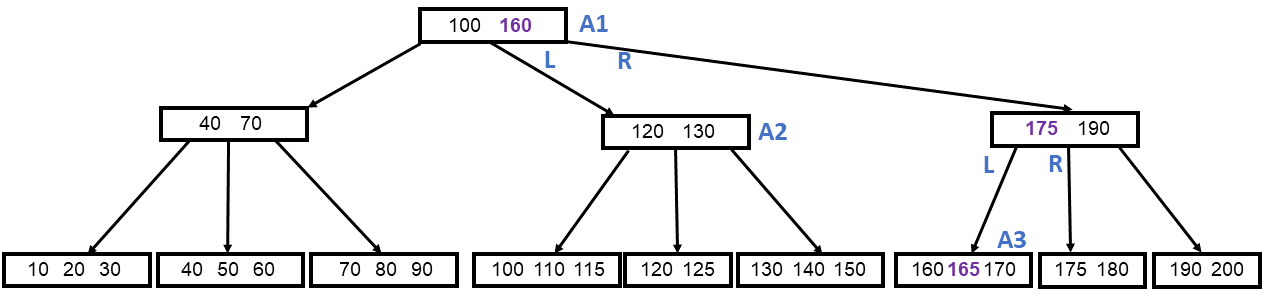}
	\caption{B-tree After Insert of 165}
	\label{exampleinsert}
\end{figure}

Figures \ref{exampletree} and \ref{exampleinsert} illustrate the insertion and splitting process. Both interior and leaf nodes have a maximum of 4 keys. The value inserted is {\bf 165}. During searching for the insert leaf, the algorithm tracks the insert path from root to leaf, which in this example are nodes with ids {\bf A1}, {\bf A2}, and {\bf A3}. The leaf node is full so a split occurs of the leaf. The insert location is at index 1. The first step buffers the mid record 175 at index~2. Record 170 is shifted down a location and record 165 is inserted at index 1. The left page is written to storage. The mid record 175 is then copied to the start of the page and the remaining records in the block copied after it. The right page is written to storage. The insertion at the next level up (A2) is 175 with the left and right pointers to the pages just written. This block also overflows. The mid record is 160. The left page contains only 120 and 130 and requires no record copying. The right page begins with the inserted record 175 and the remaining records in the block. There is space in the root (A1) which is updated with the new key (160) and left and right pointers.

Given a B-tree of height $H$, the insert operation requires $H$ reads when searching for the leaf node, and a maximum of $2*H+1$ writes if the split operation splits nodes all the way to the root. For most use cases, the height is less than 5 and splits are infrequent. Only 1 write buffer is required regardless how many splits are performed. This is a significant improvement compared to previous implementations that would use two buffers for each split level (i.e. up to $2*H$ buffers). Thus, the insert performance is very efficient. Given that only one buffer is used for writes, the B-tree can use the other buffer(s) to concurrently perform queries without affecting the updates.

Searching for a key requires $H$ reads. Multiple readers can use the B-tree structure concurrently. Deletion is rare for embedded use cases, and nodes that are under full are not merged together. This reduces the writes performed.

\section{Experimental Evaluation}

Experiments test the B-tree implementation performance on an Arduino MEGA 2560 that uses a 8-bit AVR ATmega2560 microcontroller and has 256~KB of flash program memory, 8~KB of SRAM, 4~KB EEPROM, and supports clock speeds up to 16 MHz. Storage was on a 8 GB microSD card attached with an Arduino Ethernet shield. The page size was 512 bytes. There were no hardware buffers on the SD card accessible to the algorithm, so all page buffering was in RAM. With a page size of 512 bytes, the practical limit of the number of memory buffers, $M$, on the device was $M=8$ (4 KB) as the rest of the RAM was used for other functions. The memory usage for the B-tree was 1.5 KB (for $M=2$) and the code space usage was 10 KB (about 1500 lines of code).

The SD card sequential read performance was 345 pages/second (172 KB/second), and sequential write performance was 175 pages/second (88 KB/second). Random read performance was also 345 pages/second (172 KB/second). The write-to-read ratio is 1.95. On this hardware, data transmission on the bus limited performance as much as the raw SD card performance characteristics.

The experimental data set was chosen to be representative of real-world sensor data sets that often consist of an integer key or timestamp and some collected data. The data records were 16 bytes with a 4 byte integer key. With a page size of 512 bytes, the number of records per page was 31 as each page had a page header that consumed some space. The page header stores the page id and the count of the number of records in the page. The number of interior records per page was 62. Each interior record is 8 bytes consisting of a 4 byte integer key and a 4 byte page id. Data sets were randomly generated. The results are the average of three runs. Given that there are no previous B-tree implementations that run on the memory-constrained hardware, the experiments are designed to demonstrate how the B-tree implementation can be used efficiently for practical, real-world data collection use cases.

\subsection{Insert and Query Performance}

\begin{figure}[htp]
	\centering
	\includegraphics[width=4.5in]{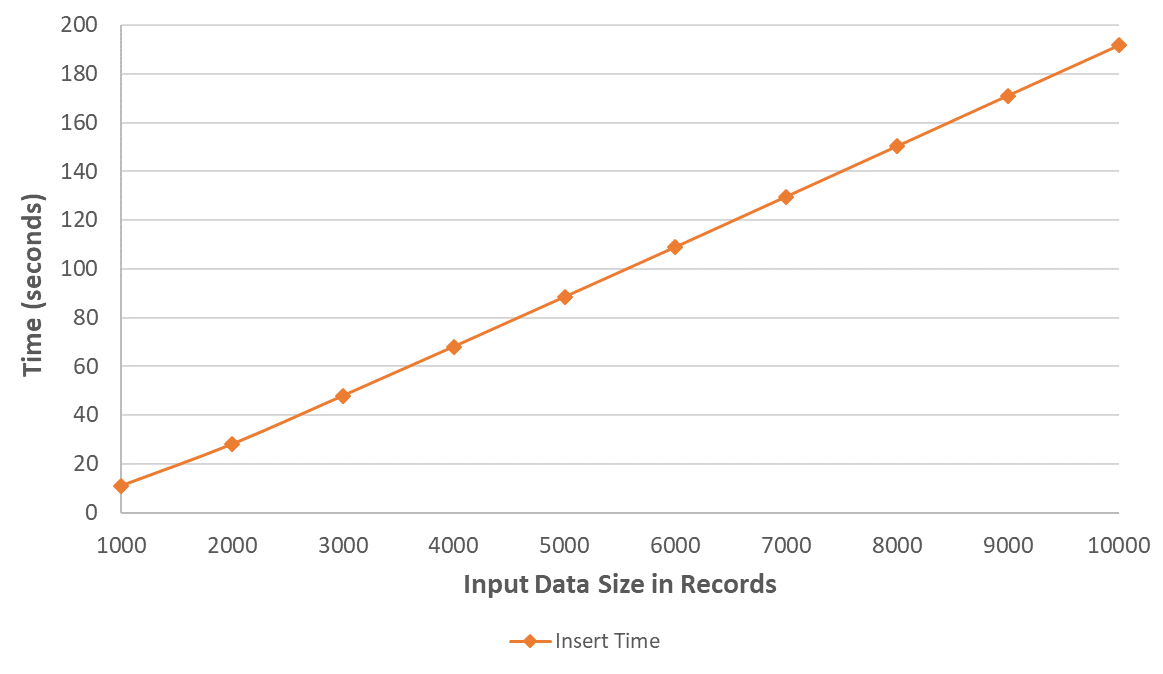}
	\caption{B-tree Insert Time}
	\label{inserttime}
\end{figure}

Figure \ref{inserttime} shows the time to insert up to 10,000 records (160 KB) for $M=2$. At 10,000 records the data size is 80 times larger than the RAM available. In server systems with more memory available, this ratio is between 2 to 10. The height of the tree is 3 by 10,000 records, and the overall time is linear in the number of records inserted.

\begin{figure}[htp]
	\centering
	\includegraphics[width=4.5in]{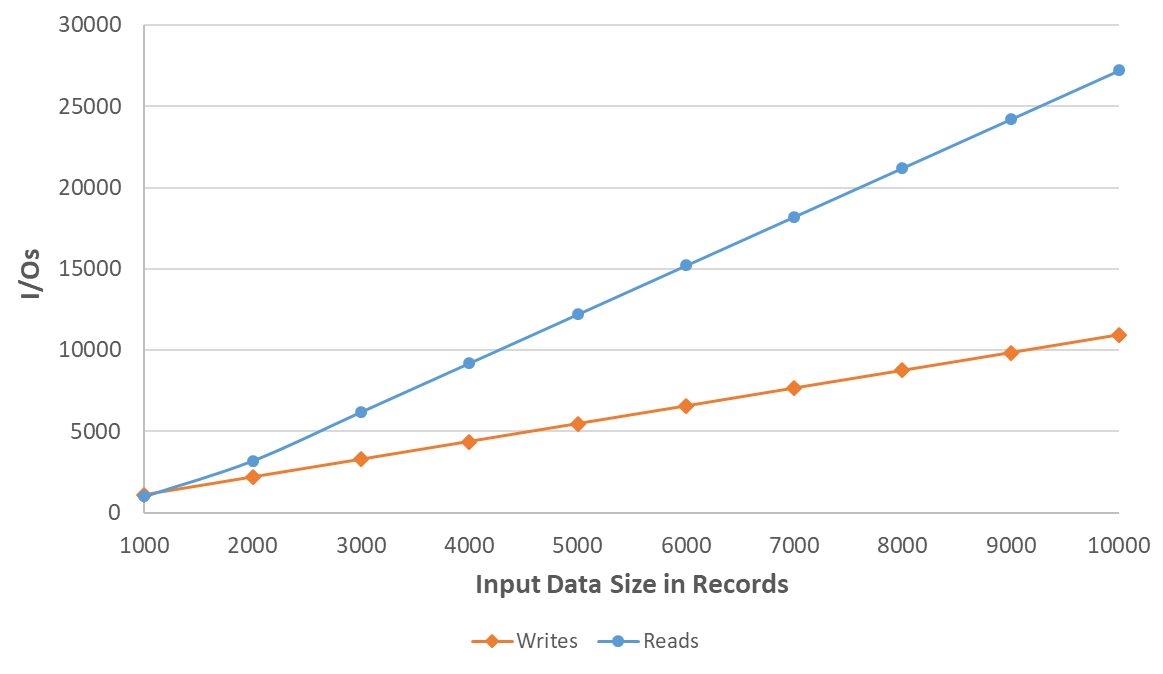}
	\caption{B-tree Insert I/O}
	\label{insertio}
\end{figure}

Figure \ref{insertio} shows the number of read and write I/Os for the insert. The performance is extremely consistent and exactly follows the expected theoretical performance formula. Each insert requires $H$ page reads with $H$ being 3 starting at 3000 records. Every insert requires at least 1 write. Inserting 10,000 records requires 10,940 writes which is only 9.4\% more writes than the minimum of 10,000 writes required if every record is written to storage immediately.

\begin{figure}[htp]
	\centering
	\includegraphics[width=4.5in]{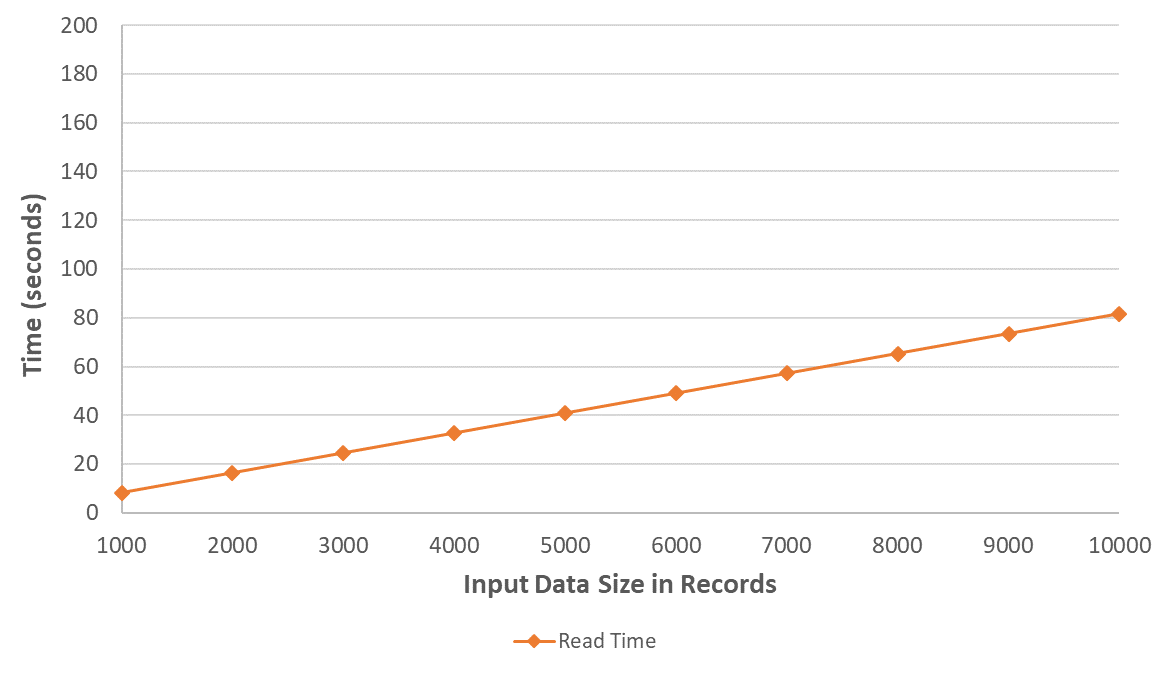}
	\caption{B-tree Query Time}
	\label{querytime}
\end{figure}

\begin{figure}[htp]
	\centering
	\includegraphics[width=4.5in]{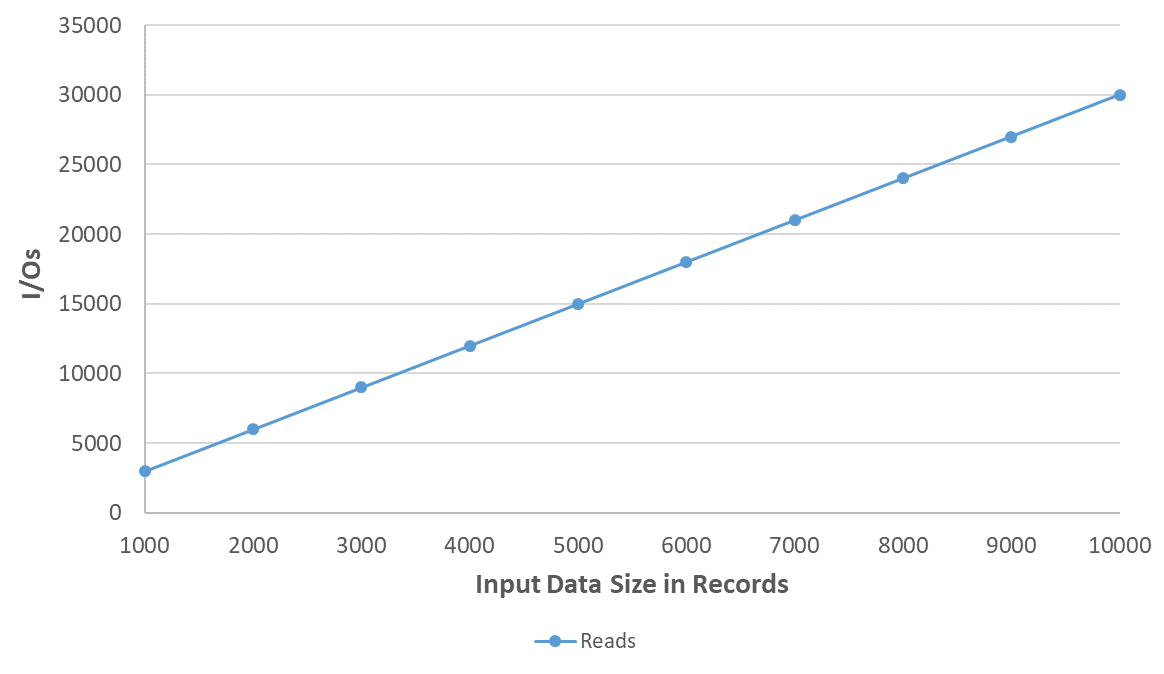}
	\caption{B-tree Query I/O}
	\label{queryio}
\end{figure}

Figure \ref{querytime} shows the time to query 10,000 random records (160 KB) for $M=2$ from a B-tree with 10,000 records. Figure \ref{queryio} shows the number of reads. Each record query requires $H=3$ page reads, and the performance is linear and stable. In comparison to sequential files which are often used, retrieving a record in a sequential file requires $N/2$ reads where $N$ is the number of data pages. For this experiment, this results in 161 page reads which is over 50 times larger.

The results show that the implementation performs inserts and queries efficiently with a minimal amount of memory.

\subsection{Memory Buffering}

The implementation is designed to use additional memory provided beyond the minimum two buffers. If given an additional buffer, then that buffer is always used to buffer the root node as this results in the greatest performance benefit. Any additional buffers beyond three are allocated using a least-recently used (LRU) algorithm. With random inserts and queries, besides the root, the pages requested are random and the buffer hit rate is low. 

\begin{figure}[htp]
	\centering
	\includegraphics[width=4.5in]{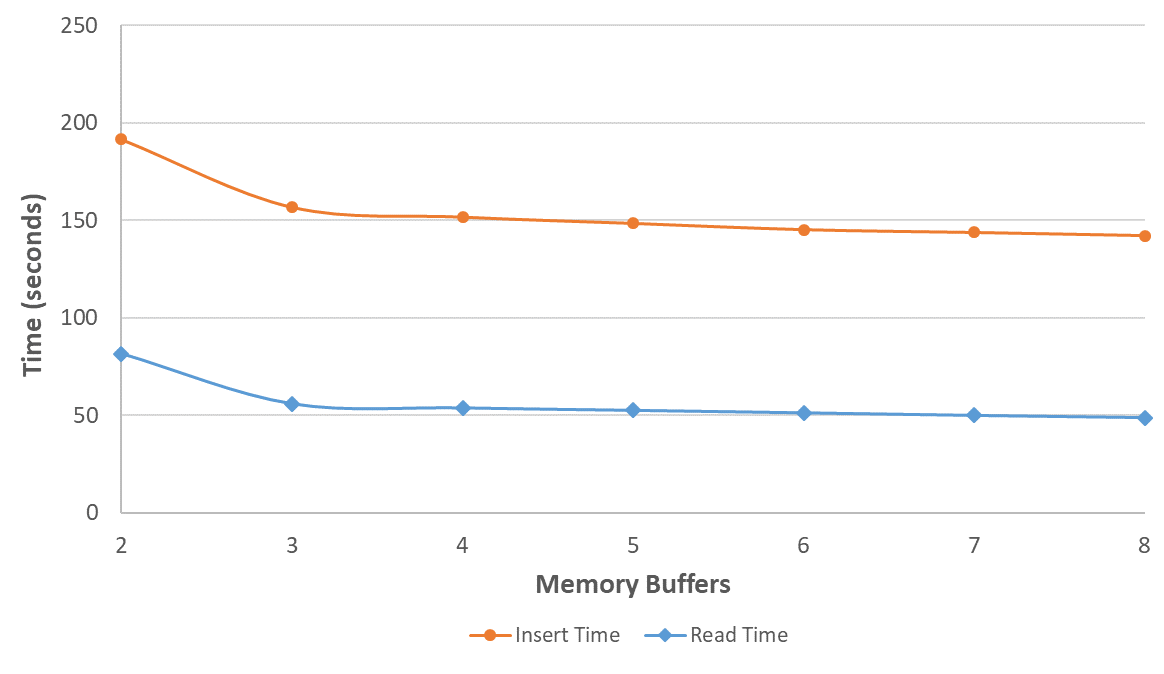}
	\caption{B-tree Time Performance with Increasing Memory}
	\label{increase_m_time}
\end{figure}

\begin{figure}[htp]
	\centering
	\includegraphics[width=4.5in]{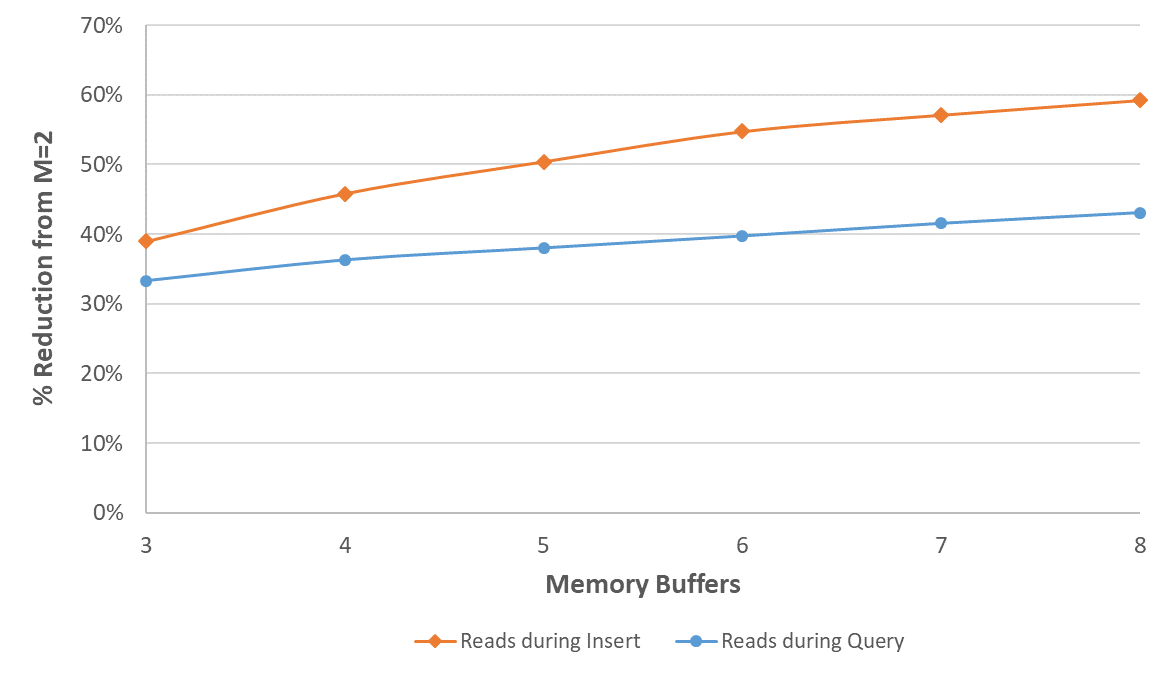}
	\caption{B-tree I/O Performance with Increasing Memory}
	\label{increase_m_io}
\end{figure}

Figure \ref{increase_m_time} shows the time to perform insert of 10,000 records and querying of the 10,000 records in random order for various values of $M$. Figure \ref{increase_m_io} shows the reduction in reads during insert and query as $M$ increases. The experiments show that additional memory has advantages, but the jump to $M=3$ is the biggest benefit. Adding a third buffer increases the buffer hit rate to 33\% for queries and almost 40\% for inserts, which makes sense given that the root is used in every query and insert operation. The third buffer decreases the insert time by 20\% and the query time by 30\%. Additional buffers slightly decrease reads, but the buffer hit rate improvement is low as the pages accessed are random. Thus, designers with limited memory are typically deciding between $M=2$ or $M=3$ and additional memory is less useful.

\subsection{Example Use Case}

The previous experiment results demonstrate the efficient performance of the B-tree implementation. In practice, it is important that the B-tree is deployable in real-world sensor data collection environments. An example use case is collecting environmental data for sustainable agriculture and environmental analysis. In \cite{irrigation}, soil moisture data was collected to determine optimal irrigation patterns to reduce water usage. The data was sampled every minute, stored on the device, and used to control the irrigation system. The system was effective but required custom implementation of data structures for data storage.

A key metric is the time to perform a single insert as this limits the frequency that the sensor can sample data. The B-tree implementation with $M=2$ requires 20~ms per insert and 8 ms per query. For $M=3$, the time is 15 ms per insert and 5.5 ms per query. Given this performance, a deployment using the B-tree can sample up to 50 times/second. In practice, most use cases sample every minute or 15 minutes, and the percent time spent on data storage is less than 0.03\%.

Overall, the B-tree implementation provides an efficient data structure for data storage on embedded devices and supports practical use cases of data collection with minimal memory and power usage. The code is available on GitHub (https://github.com/ubco-db) with an open source license to allow it to be used by the community.

\section{Conclusions and Future Work}

Indexing on embedded systems requires implementations that are memory efficient otherwise they are not practically useful for devices with limited memory and computational capabilities. The B-tree implementation uses the minimum possible two buffers and provides high performance for insert and queries. Using a third buffer significantly improves performance. Experimental results demonstrate the efficiency of the implementation and its practical benefit for a typical sensor monitoring application. 

Future work will perform further experiments on other embedded hardware platforms and flash memory configurations including raw flash memory rather than using SD card storage. The B-tree index will also be deployed in an embedded database system used for environmental sensor monitoring applications.

\section{Acknowledgment}

The authors would like to thank NSERC for supporting this research.

\bibliographystyle{spmpsci}
\bibliography{refs}

\end{document}